\documentclass{emulateapj}

\newcommand{\myemail}{srazzaque@ssd5.nrl.navy.mil}

\newcommand{\be}{\begin{equation}}
\newcommand{\ee}{\end{equation}}
\newcommand{\ba}{\begin{eqnarray}}
\newcommand{\ea}{\end{eqnarray}}
\def\eps{\epsilon}
\def\veps{\varepsilon}
\def\a{\alpha}
\def\b{\beta}
\def\g{\gamma}
\def\G{\Gamma}
\def\p{\prime}

\shorttitle{Leptonic-hadronic model for GRB 090510 afterglow} 
\shortauthors{Razzaque}

\begin{document}

\title{A Leptonic-hadronic model for the afterglow of gamma-ray burst
090510}

\author{Soebur Razzaque\altaffilmark{1,2}}

\altaffiltext{1}{Space Science Division, Code 7653, U.S. Naval
Research Laboratory, 4555 Overlook Ave SW, Washington, DC 20375;
\myemail}
\altaffiltext{2}{National Research Council Research Associate}

\begin{abstract}
We model multiwavelength afterglow data from the short Gamma-Ray Burst
(GRB) 090510 using a combined leptonic-hadronic model of synchrotron
radiation from an adiabatic blast wave.  High energy, $\gtrsim
100$~MeV, emission in our model is dominated by proton-synchrotron
radiation, while electron-synchrotron radiation dominates in the X ray
and ultraviolet wavelengths.  The collimation-corrected GRB energy,
depending on the jet-break time, in this model could be as low as
$3\times 10^{51}$~erg but two orders of magnitude larger than the
absolute $\g$ ray energy.  We also calculated the opacities for
electron-positron pair production by $\g$ rays and found that TeV $\g$
rays from proton-synchrotron radiation can escape the blast wave at
early time, and their detection can provide evidence of a hadronic
emission component dominating at high energies.
\end{abstract}


\maketitle

\section{Introduction}

Gamma-Ray Burst science has entered a new era with launch of the {\em
Fermi} $\g$-ray space telescope.  The main instrument, Large Area
Telescope (LAT), is more sensitive than any previous instrument in the
20~MeV -- 300~GeV range~\citep{LAT}, whereas the {\em Fermi} Gamma-ray
Burst Monitor (GBM) observes the whole unocculted sky in the 8~keV --
40~MeV range~\citep{GBM}.  With the advent of the Burst Alert
Telescope (BAT), X-Ray Telescope (XRT) and UV-Optical Telescope (UVOT)
onboard the {\em Swift} satellite~\citep{Swift} it is now possible to
obtain simultaneous multiwavelength data in the optical to multi-GeV
$\g$-ray energy range from GRBs.

GRB 090510 is the first GRB to provide data from simultaneous
observtions by {\em Fermi}~\citep{grb090510_Fermi} and {\em
Swift}~\citep{grb090510_Swift}, as well as by a couple of other
satellites.  At a redshift $z=0.903\pm
0.003$~\citep{grb090510_redshift} the isotropic-equivalent $\g$-ray
energy release from this short GRB ($T_{90} \lesssim 2$~s) is
$E_{\g,\rm iso} = (1.08\pm 0.06)\times 10^{53}$~erg with a fluence of
$(5.03\pm 0.25)\times 10^{-5}$~erg~cm$^{-2}$ in the 10~keV -- 30~GeV
range~\citep{grb090510_Fermi}.  {\em Fermi} LAT detected long-lived
emission up to $\sim 200$~s after trigger ($T_0 =$~00:23:00~UT, 2009
May 10) in the $\gtrsim 100$~MeV range.  While such high-energy
emission, that is temporally extended beyond the keV -- MeV emission,
was first detected in GRB 940217 by the {\em Compton Gamma-Ray
Observatory}~\citep{hur94}, this feature is common to most GRBs
detected with {\em Fermi} LAT.  {\em Swift} XRT and UVOT collected
data from GRB 090510 between $\sim T_0+97$~s and $T_0+1.9$~ks before
an Earth Occultation (EO), and again after
$T_0+5.1$~ks~\citep{grb090510_Swift}. {\em Swift} BAT collected most
data within $T_{90} = 0.3\pm 0.1$~s (15 -- 350~keV), and sparsely
between $T_0+0.4$~s and $\sim T_0+100$~s~\citep{grb090510_Swift}.

Smooth temporal evolution of the flux, $F\propto t^{-\a}$, of the
long-lived emission in {\em Fermi} LAT ($\a_\g = 1.38\pm 0.07$), and
{\em Swift} XRT ($\a_{X,1} = 0.74\pm 0.03$ before EO) and UVOT
($\a_{O,1} = -0.50^{+0.11}_{-0.13}$ before EO) observations strongly
suggest an afterglow origin~\citep{grb090510_afterglow}.  Synchrotron
radiation by shock-accelerated electrons in a decelerating GRB blast
wave~\citep{mr97,spn98} have successfully explained much of the
broadband afterglow data at radio, optical, and X-ray frequencies in
the pre-{\em Fermi} era.  However fitting combined {\em Fermi} and
{\em Swift} data from GRB 090510 with simple $e$-synchrotron model
results in unusual parameter values, and most importantly it is
difficult to reconcile the $F_\nu \propto t^{-\alpha} \nu^{-\beta}$
temporal relations~\citep{grb090510_afterglow}.  More complex
scenarios have been proposed to model GRB 090510 data such as a
radiative fireball in an $e^\pm$ pair dominated
environment~\citep{ggn09,ggnc09}, adiabatic fireball in a low density
medium and small magnetic field~\citep{kb09a,kb09b,gmxf09}, and two
component jet~\citep{cgp09}.

Here we present a combined leptonic- and hadronic- afterglow model to
fit multiwavelength data from GRB 090510.  Inclusion of ion
acceleration and radiation in the GRB blast wave is a natural and
simple extension of the $e$-synchrotron blast wave model, and has been
discussed by a number of authors~\citep{bd98,tot98a,zm01,wan09,rdf10}.
We show that LAT emission in the $\gtrsim 100$~MeV range is dominated
by synchrotron radiation from protons accelerated in the external
forward shock of a decelerating blast wave.  (Note that \citet{rdf10}
considered proton-synchrotron radiation from a coasting blast wave.)
The XRT and UVOT light curves can be reasonably reproduced by
synchrotron radiation from electrons accelerated in the same external
forward shock.  We present the $e$- and ion- synchrotron afterglow
model in Sec.\ 2, compare this model with GRB 090510 afterglow data in
Sec.\ 3, and discuss our results in Sec.\ 4.

\section{Synchrotron afterglow model}
\label{sec:model}

With a coasting bulk Lorentz factor of $\G_0 =10^3\G_{3}$ and an
isotropic-equivalent kinetic energy $E_{\rm k, iso} =
10^{55}E_{55}$~erg, the deceleration time scale for an adiabatic blast
wave in a medium of uniform density $n$~cm$^{-3}$ is~\citep{bm76,spn98}
\be
t_{\rm dec} \approx 1.9 ~(1+z) (E_{55}/n)^{1/3} \G_{3}^{-8/3}~{\rm s}.
\label{dec_time}
\ee
Later the bulk Lorentz factor evolves as 
\be
\G \approx 763 ~(1+z)^{3/8} (E_{55}/n)^{1/8} t_s^{-3/8},
\label{bulk_G}
\ee
where $t_s~(>t_{\rm dec})$ is measured in seconds.  At the
deceleration time $t=t_{\rm dec}$, $\Gamma \approx \Gamma_0/2^{3/4}$.
The radius of the blast wave, $R = 4\G^2 c t$, is given by
\be
R \approx 1.4\times 10^{17} (1+z)^{-1/4} 
(E_{55} t_s/n)^{1/4}~{\rm cm}.
\label{BW_radius}
\ee
The jet-break time at which $\G \approx \theta_0^{-1}$ \citep{sph99},
where $\theta_{0} = 0.1\theta_{-1}$ is the jet opening angle, is given
by
\be
t_{\rm jet} \approx 10^5~(1+z) (E_{55}/n)^{1/3} \theta_{-1}^{8/3}
~{\rm s}.
\label{jet_brk}
\ee

The fractions of energy injected in a forward shock~\citep{bm76} that
channel into electrons\footnote{Refererring to both electrons and
positrons.} and into ions can be calculated from their
shock-accelerated spectra.  We assume an electron injection
spectrum\footnote{Comoving frame variables are denoted with primes.}
$n^\p_e (\g^\p_e) \propto \g^{\p -k}_e$ for $\g^\p_{m,e} \le \g^\p_e
\le \g^\p_{sat,e}$.  Here $\g^\p_{m,e} = \eta_e (m_p/m_e)\G (t)$ and
$\g^\p_{sat,e}$ are the minimum and saturation Lorentz factors,
respectively, for the electrons. In case of ions, we assume an
injection spectrum $n^\p_A (\g^\p_A) \propto \g^{\p -k_1}_A$ for
$\G(t) \le \g^\p_A \le \g^\p_{m,A}$ and $n^\p_A (\g^\p_A) \propto
\g^{\p -k_2}_A$ for $\g^\p_{m,A} \le \g^\p_A \le \g^\p_{sat,A}$.  Here
$\g^\p_{m,A} = \eta_A\G (t)$ is a break in the spectrum and
$\g^\p_{sat,A}$ is the saturation ion Lorentz factor.  The fraction of
shock energy carried by the electrons is
\be
\eps_e \simeq \xi_e\eta_e \frac{k-1}{k-2} ~
\frac{1-(\g^\p_{m,e}/\g^\p_{sat,e})^{k-2}}
{1-(\g^\p_{m,e}/\g^\p_{sat,e})^{k-1}} ;~ k \ne 2
\label{epsilon_e}
\ee
and that by the ions is
\be
\eps_A \simeq \xi_A\eta_A \frac{k_1-1}{k_1-2} ~
\frac{\frac{k_1-2}{k_2-2} +\eta_A^{k_1-2} -1}
{\frac{k_1-1}{k_2-1} +\eta_A^{k_1-1} -1} ;~ 
\matrix{k_1\ne 1 \cr k_2 \gg 2 .}
\label{epsilon_A}
\ee
Here $\xi_e$ and $\xi_A$ are the number fractions of electrons and
ions that are accelerated by the shock, respectively, with an equal
pre-shock number density $n \equiv n_e = n_A$.

A fraction $\eps_B$ of the shock energy is assumed to generate
magnetic field, and the magnetic field behind the forward shock is
given by~\citep{spn98}
\be
B^\p \approx 297 ~(1+z)^{3/8}\eps_B^{1/2} 
(E_{55}n^3)^{1/8} t_s^{-3/8} ~{\rm G}.
\label{B_field}
\ee
In the regime of our interest $\eps_B \gg \eps_e$, the Compton
parameter $Y = [-1+\sqrt{1 +4\eps_e/\eps_B}]/2 \to 0$ and the energy
loss by the electrons is dominated by synchrotron
radiation~\citep{se01}.

The saturation Lorentz factor for electrons is calculated by equating
the acceleration time to the synchrotron cooling time in the $B^\p$
field (eq.~[\ref{B_field}]) as
\be
\g^\p_{sat,e} \approx \frac{ 6.8\times 10^{6} ~t_s^{3/16}}
{(1+z)^{3/16} \phi_e^{1/2}\eps_B^{1/4} (E_{55} n^3)^{1/16}}.
\label{e_sat_Lorentz}
\ee
Here $\phi_e^{-1}$ is the acceleration efficiency for electrons.  The
cooling Lorentz factor, found by equating the synchrotron cooling time
to the dynamic time $t^\p_{dyn} = t\G/(1+z)$, is given by
\be
\g^\p_{c,e} \approx 11.5 ~(1+z)^{-1/8}\eps_B^{-1} 
(E_{55}^3n^5)^{-1/8} t_s^{1/8}.
\label{e_c_Lorentz}
\ee

For ions, of atomic mass $A$ and charge $Z$, the saturation Lorentz
factor is calculated by equating the acceleration time to the shorter
of the dynamic time and synchrotron cooling time as
\ba
\g^\p_{sat,A} &\approx & 
\frac{2.2\times 10^9 (Z/A) \eps_B^{1/2} (E_{55}n)^{1/4} t_s^{1/4}}
{(1+z)^{1/4} \phi_A} ;~ t< t_{d,A}
\nonumber \\ \g^\p_{sat,A} &\approx & 
\frac{ 1.2\times 10^{10} (A/Z^{3/2}) t_s^{3/16}}
{(1+z)^{3/16} \phi_A^{1/2}\eps_B^{1/4} (E_{55} n^3)^{1/16}}
;~ t\ge t_{d,A}.\,\;\;\;\;\;
\label{A_sat_Lorentz}
\ea
The transition takes place at 
\be
t_{d,A} \approx 1.4\times 10^{12} (1+z) \phi_A^8 \eps_B^{-12}
n^{-7} E_{55}^{-5}~ {\rm s}.
\label{A_sat_time}
\ee
Note that the cooling Lorentz factor for ions $\g^\p_{c,A} =
(A^3/Z^4)(m_p/m_e)^3 \g^\p_{c,e}$ can be larger than the saturation
Lorentz factor (eq.~[\ref{A_sat_Lorentz}]).

To calculate synchrotron spectra at different epoch and light curves
at different frequencies arising from a forward shock, it is
sufficient to calculate different spectral break frequencies and flux
normalization along with their time
evolution~\citep[e.g.][]{spn98,cl00,pk00,gs02}.

The characteristic synchrotron frequencies for the electrons with the
minimum, saturation and cooling Lorentz factors respectively are given
by
\ba
&& h\nu_{m,e} \approx 7.7
(1+z)^{1/2} \eta_e^2 (\eps_B E_{55})^{1/2} t_s^{-3/2} ~{\rm GeV},
\nonumber \\
&& h\nu_{sat,e} \approx 180
(1+z)^{-5/8} \phi_e^{-1} (E_{55}/n)^{1/8} t_s^{-3/8} 
~{\rm GeV},
\nonumber \\
&& h\nu_{c,e} \approx 
0.5 (1+z)^{-1/2} \eps_B^{-3/2} (E_{55}n^2)^{-1/2} t_s^{-1/2} 
~{\rm eV}.~~~~ 
\label{e_m_sat_c_frq}
\ea
A transition from the fast-cooling ($\nu_{c,e} < \nu_{m,e}$) to
slow-cooling ($\nu_{c,e} > \nu_{m,e}$) takes place at
\be
t_{0,e} \approx 1.5\times 10^{10} (1+z)(\eps_B\eta_e)^2 nE_{55} 
~{\rm s}.
\label{fast_slow}
\ee
In both the fast- and slow- cooling cases the maximum $e$-synchrotron
flux is given by~\citep[e.g.][]{spn98}
\be
F_{\nu, e}^{\rm max} \approx 52~
(1+z)^{-1} \xi_e (\eps_B n)^{1/2} d_{28}^{-2} E_{55} ~{\rm Jy}.
\label{e_max_F}
\ee
Here $d_{28}/(10^{28}~{\rm cm})$ is the luminosity distance.  Note
that the synchrotron self-absorption frequency is in the radio
band~\citep[e.g.][]{pk00} and we ignore that while modeling
optical to $>$ GeV data.

The synchrotron break frequencies for the ions of minimum and cooling
Lorentz factors can be expressed as scaling relations to the
corresponding break frequencies for electrons as
\ba
\nu_{m,A} &=& Z(\eta_A/\eta_e)^2 (m_e/m_p)^3 \nu_{m,e},
\nonumber \\
\nu_{c,A} &=& (A^6/Z^7) (m_p/m_e)^5 \nu_{c,e},
\label{A_sync_brk_m_c}
\ea
and for the ions of saturation Lorentz factor
(eq.~[\ref{A_sat_Lorentz}]) as
\ba
&& h\nu_{sat,A} \approx  \frac{10~ \eps_B^{3/2} (nE_{55})^{3/4}}
{(1+z)^{3/4} \phi_A^{2} t_s^{1/4}}~{\rm TeV} ;~ t<t_{d,A}
\nonumber \\
&& \nu_{sat,A} = (A/Z)^2 (m_p/m_e)(\phi_e/\phi_A) \nu_{sat,e}
;~ t\ge t_{d,A}. ~~~~~~
\label{ion_sync_break}
\ea
Note that the ion-synchrotron spectrum is always in the slow-cooling
regime ($\nu_{c,A} > \nu_{m,A}$) as opposed to the $e$-synchrotron
spectrum which can be in the fast-cooling regime early and changes to
the slow-cooling regime later.  The maximum ion synchrotron flux is
\ba
F_{\nu,A}^{\rm max} \simeq \frac{k_1 -1}
{\frac{k_1-1}{k_2-1} +\eta_A^{k_1-1} -1}
\frac{\xi_A}{\xi_e}
\frac{Z^3}{A^2} \frac{m_e}{m_p} F_{\nu,e}^{\rm max},
\label{A_max_F}
\ea
for $k_1 \ne 1$ and $k_2>2$.

\section{Modeling GRB 090510 afterglow data}

\begin{figure}
\includegraphics[width=3.3in]{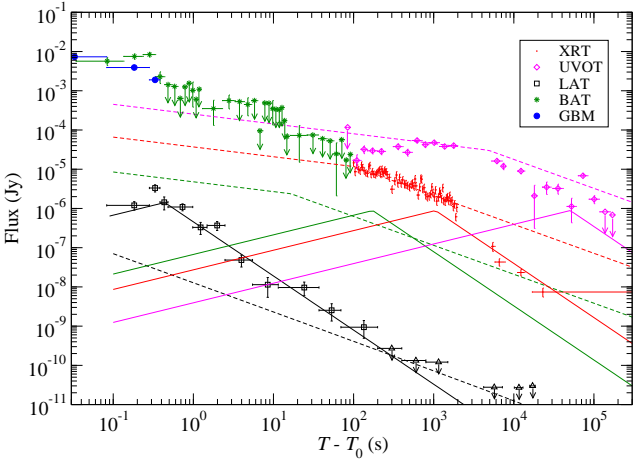}
\caption{
Modeling of GRB 090510 light curves with proton-synchrotron (solid
lines) and electron-synchrotron (dashed lines) radiation from an
adiabatic blast wave decelerating in a uniform density medium.  The
model light curves are computed at 100~MeV (black), 15~keV (green),
1~keV (red) and 3~eV (magenta) to be compared with the LAT, BAT, XRT
and UVOT data points, respectively.  The onset of the model light
curves is at $\lesssim T_0+0.3$~s for $\G_0 \gtrsim 2400$ for a
surounding medium density of $n= 3$~cm$^{-3}$.  The other model
parameters are $E_{\rm k,iso} = 2\times 10^{55}$~erg, $\eps_e \approx
10^{-4}$, $\eps_p \approx 0.5$, $\eps_B = 0.3$.  See main text for
more details.}
\label{fig:synchrotron}
\end{figure}

Figure~\ref{fig:synchrotron} shows light curves at different energies
from the combined leptonic-hadronic model of a decelerating adiabatic
blast wave in constant density medium.  With an initial $\G_0 \gtrsim
2400$, the blast wave decelerates at $\lesssim 0.3$~s
(eq.~[\ref{dec_time}]) for the parameters used here: $E_{55}\approx
2$, $n\approx 3$~cm$^{-3}$.

We model LAT emission in the $\approx$~100~MeV -- 4~GeV range in
Fig.~\ref{fig:synchrotron} from $p$-synchrotron radiation ($A=1$) for
which $\nu_{m,p} < \nu_{p} < \nu_{c,p}$.  In order to reproduce the
flux decay index $\a_\g = 1.38\pm 0.07$ in this range for a
slow-cooling spectrum, one requires $k_2 = (4/3)\a_\g +1 = 2.84\pm
0.09$.  The corresponding spectral index is $\beta_\g = (k_2 -1)/2 =
(2/3)\a_\g = 0.92\pm 0.05$.  For $\nu_{p} < \nu_{m,p} < \nu_{c,p}$,
the $p$-synchrotron flux scales as $F_\nu \propto t^{1/2} \nu^{1/3}$.
A constraint on $h\nu_{m,p} \lesssim 100$~MeV as early as $\approx
T_0+0.4$~s requires that $\eta_p \lesssim 5\times 10^3$
(eq.~[\ref{A_sync_brk_m_c}]) with $\eps_B \approx 0.3$.  These
requirements together with the flux level (eq.~[\ref{A_max_F}]) needed
to reproduce LAT data constrain the $p$-synchrotron radiation
component.  The fraction of jet energy needed in shock-accelerated
protons (eq.~[\ref{epsilon_A}]) is $\eps_p \approx 0.5$ for $k_1
\lesssim 0$ and $\xi_p \approx 10^{-4}$.  The rise of the $\gtrsim
100$~MeV LAT flux at $T\lesssim T_0+0.4$~s could be consistent with
the $t^2$ rise before the blast wave enters the self-similar
regime~\citep{sari97}.  The $p$-synchrotron flux in the optical to
X-ray is much below the XRT and UVOT data.

We reproduce the XRT light curve, averaged over 0.3 -- 10~keV range,
with decay index $\a_{X,1} = 0.74\pm 0.03$ before the EO at $T\approx
T_0 + 1.43$~ks as from $e$-synchrotron radiation.  The required
electron index is $k = (4/3)\a_{X,1}+2/3 = 1.65\pm 0.04$ for $\nu_e
>\nu_{m,e} > \nu_{c,e}$ in the fast-cooling case, which is valid for a
time $T\lesssim T_0+2\times 10^{6}$~s (eq.~[\ref{fast_slow}]).
Note that the spectral index $\b_{X,1} = k/2 = (2\a_{X,1}+1)/3 =
0.83\pm 0.02$, is close to that of $\b_\g$ from $p$-synchrotron
radiation.  In order to produce $h\nu_{m,e} \lesssim 1$~keV at the
beginning of XRT observation at $T\approx T_0+100$~s, we require
$\eta_e \lesssim 20(m_e/m_p)$ (eq.~[\ref{e_m_sat_c_frq}]).
Together with parameter $\xi_e \approx 5\times 10^{-4}$ required to
produce the XRT flux level, we calculate the fraction of jet energy in
electrons to be $\eps_e \approx 10^{-4}$ (eq.~[\ref{epsilon_e}])
with $\phi_e = 1$.

Electron-synchrotron flux in the UVOT range is in the frequency range
$\nu_{c,e} < \nu_{e} < \nu_{m,e}$ and scales as $F_\nu \propto
t^{-1/4} \nu^{-1/2}$.  Although the observed flux fitted with
$\a_{O,1} = -0.50^{+0.11}_{-0.13}$~\citep{grb090510_afterglow} is
different, we note that the expected value of $\a = 1/4$ is consistent
with UVOT data in the $T-T_0 \approx 600$~s -- 1.43~ks interval, and
with the upper limit at $T\approx T_0 +90$~s.

It is clear that the observed X-ray and UVOT flux decay indices
$\a_{X,2} = 2.18\pm 0.1$ and $\a_{O,2} = 1.13^{+0.11}_{-0.10}$,
respectively, after the EO at $T\gtrsim T_0+5.1$~ks
~\citep{grb090510_afterglow} are softer than the $e$-synchrotron
emission. If the jet break takes place in between $T-T_0 \approx
1.4$~ks -- 5.1~ks, then the expected decay index for $\nu_{e} >
\nu_{m,e} > \nu_{c,e}$ is $\propto t^{-k}$ which is intermediate
between $\a_{O,2}$ and $\a_{X,2}$ since $k=1.65$.  Because of the
idealized nature of the afterglow model and evolution of the blast
wave during and after the jet break~\citep[e.g.][]{sph99}, the
observed flux steepening after the EO could still be due to a jet
break.  For $1.4~{\rm ks} \lesssim t_{\rm jet}-T_0 \lesssim 5.1~{\rm
ks}$ the jet opening angle is $0.16 \lesssim \theta_{-1} \lesssim
0.26$ (eq.~[\ref{jet_brk}]).  If the jet-break takes place at $T
\gtrsim T_0+100$~ks, then $\theta_{-1} \gtrsim 0.8$.

As shown in Fig.~\ref{fig:synchrotron} the BAT flux is quite noisy and
can not be reproduced by either $e$- or $p$- synchrotron emission.  A
similar conclusion was drawn by~\citet{grb090510_afterglow} based
solely on $e$-synchrotron afterglow model.  Sporadic emission in the
BAT range could be due to central engine activity, working
intermittenly at a much reduced emission level than the initial
outburst.

During the early deceleration phase, the soft photon density in the
GRB blast wave may be large enough to induce $\g\g\to e^+e^-$ pair
production and photohadronic ($p\g$) interactions by protons, and
subsequent cascade formation.  The target photon density can be
calculated as $n^\p_\g (\veps^\p) = 2d_L^2 (1+z)F_\nu/(R^2
c\G\veps^\p)$ from the synchrotron flux, where $\veps^\p \equiv
h\nu^\p = h\nu(1+z)/\G$.

\begin{figure}
\includegraphics[width=3.3in]{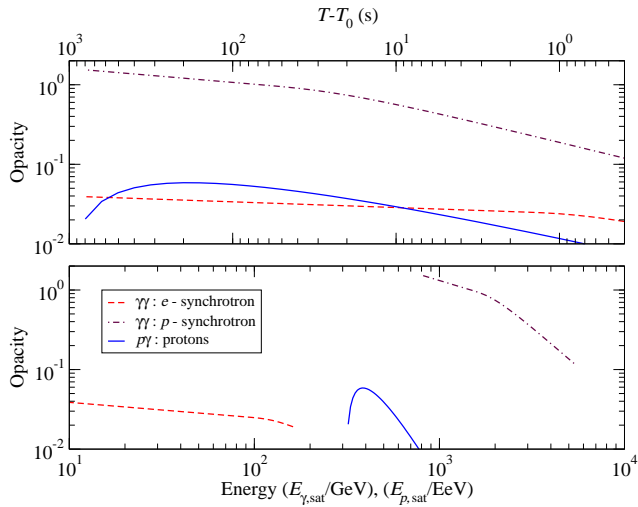}
\caption{
Opacities for the electron-positron pair production at the saturation
energies of the electron-synchrotron (dashed lines) and
proton-synchrotron (dot-dashed lines) emission.  Also shown is the
opacity for photopion ($p\g$) interaction (solid lines).  The
horizontal axis in the {\em top panel} corresponds to the opacities at
diferent time, and in the {\em bottom panel} it corresponds to the
saturation energies reached within the same time interval for each
lines.  We used the same parameters as used in
Fig.~\ref{fig:synchrotron} to calculate the opacities.
}
\label{fig:opacity}
\end{figure}

We calculate the $\g\g$ pair production and $p\g$ pion production
opacities from their respective cooling time scales and the dynamic
time scale for the decelerating blast wave model of GRB 090510. The
opacities for the $\g$ rays with saturation energies, both from the
$e$- and $p$- synchrotron emission, and for the protons with
saturation energies in Fig.~\ref{fig:opacity}.  The top panel shows
the time dependence (from right to left) of the opacities at the
saturation energy reached at that time.  The bottom panel shows the
opacities vs.\ the saturation energies reached within the same time
interval.  Thus the whole time interval of the top panel is squeezed
to fit into each of the curves in the bottom panel.

The saturation energies for the $e$-synchrotron $\g$ rays scale with
time as $h\nu_{sat,e} \approx 115~\phi_e^{-1} t_s^{-3/8}$~GeV
(eq.~[\ref{e_m_sat_c_frq}]) for the same model parameters used in
Fig.~\ref{fig:synchrotron}.  For the $p$-synchrotron, the saturation
$\g$-ray energy is $h\nu_{sat,p}\approx 4.2~\phi_{p}^{-2}t^{-1/4}$~TeV
(eq.~[\ref{ion_sync_break}]) for $T-T_0 \lesssim t_{d,p}\approx
3.7\times 10^{13} \phi_{p}^8$~s~ (eq.~[\ref{A_sat_time}]).  Thus the
$\g$-ray saturation energies decrease with time while the opacities
increase due to a flux increase of the target photons.  However the
opacities are small to initiate a substantial $e^+e^-$ pair cascade
and accompanying radiation.  The same is true for photopion
cascade~\citep[see, however,][]{agm09}, although a small fraction of
protons above $E_p \gtrsim 300$~EeV ($\phi_{p}=1$) can escape as
cosmic rays by converting to neutrons. It is interesting to note that
the saturation proton energy decreases with time as $E_{sat,p}\approx
741 \phi_p^{-1}t_s^{-1/8}$~EeV because of a decreasing bulk Lorentz
factor $\G \approx 923~t_s^{-3/8}$ (eq.~[\ref{bulk_G}]), even though
the saturation Lorentz factor increases with time as $\g^\p_{sat,p}
\propto t_s^{1/4}$ (eq.~[\ref{A_sat_Lorentz}]).

\section{Discussion and Conclusions}

We have fitted the LAT and XRT light curves from $p$- and $e$-
synchrotron emissions, respectively, before the EO from an adiabatic
blast wave in a constant density medium (Fig.~\ref{fig:synchrotron}).
The photon index for $\gtrsim 100$~MeV emission is $1+\b_\g = 1.92\pm
0.05$ for the proton index $k_2 = 2.84\pm 0.09$, and is consistent
within $1\sigma$ of the measured value of the power-law component in
the 0.9~s -- 1.0~s time interval~\citep{grb090510_Fermi}.  At earlier
time, 0.6~s -- 0.8~s, where the power-law component is significant and
harder, the agreement is within $3\sigma$.  However emission below
$\sim 20$~keV in that time interval, which we do not model, may
contribute to the hardening of the power-law component.  Our model is
compatible with photospheric emission which can dominate in the GBM
and BAT range~\citep[e.g.][]{rm06,gmxf09,twm10}.

The isotropic-equivalent $\g$-ray energy release from GRB 090510 is
2--4 orders of magnitude larger than the typical range of $10^{49}$ -
$10^{51}$ erg for short GRBs, derived from pre-{\em Fermi} era
data~\citep{n07}.  Our model requires even larger, $10^{55}$ erg, of
isotropic-equivalent energy release in the jet~\citep[see,
however,][]{fw01,ej05}.  Thus the jet of GRB 090510 must be strongly
beamed for our model to be viable.  Depending on the jet opening angle
derived in Sec.\ 3 from our model fit, the absolute jet energy is in
between $\approx (3-7)\times 10^{51}$~erg if the jet break takes place
between 1.4 -- 5.1~ks.  For a jet break at $\gtrsim 100$~ks, the
absolute energy release is $\gtrsim 6\times 10^{52}$~erg.  Note that
the inferred absolute energy of the {\em Fermi}-LAT GRB sample is
generally high, reaching $10^{53}$~erg in some cases~\citep{c10}.  We
also note that the microphysical parameter $\epsilon_e \approx
10^{-4}$ that we derive for GRB 090510 using the combined $p$- and
$e$- synchrotron model, is much lower than the typically assumed value
for short GRBs in the range $10^{-1}$ - $10^{-2}$ which is derived
from long GRB data in the pre-{\em Fermi} era using $e$-synchrotron
model only~\citep{n07}.  Finally, a confirm detection of
large-amplitude and short-time variability, synchronized to the keV -
MeV emission, in LAT data will argue against their afterglow
origin~\citep[see, e.g.,][]{dm99} as we modeled here.

A crucial test for the hadronic emission model in the afterglows of
GRBs at redshift $\lesssim 0.5$, thus avoiding absorption in the
extragalactic background light~\citep{rdf09,frd10}, may come from
ground-based TeV telescopes such as MAGIC, VERITAS, HESS and HAWC.  An
extrapolation of the LAT flux in the $\gtrsim 200$~GeV range, where
$e$-synchrotron radiation flux is negligible and internal $\g\g$
opacity is small (Fig.~\ref{fig:opacity}), should produce similar
spectra as in LAT and lower flux from $p$-synchrotron radiation.  On
the contrary, a synchrotron-self-Compton
model~\citep[e.g.][]{dcm00,zm01} is expected to produce a spectral
hardening and flux increase in the TeV range.

\vskip 10pt

I thank {\em Fermi} LAT and GBM, and {\em Swift} team members,
especially M.~De Pasquale, C.~D.~Dermer, J.~D.~Finke, N.~Gehrels, S.~
Guiriec, V.~Pelassa and F.~Piron, for providing data and useful
comments.  This work is supported by the Office of Naval Research and
NASA {\em Fermi} Cycle II GI program.


\begin{thebibliography}{}

\bibitem[Abdo et al.(2009)]{grb090902b} Abdo, A.~A. et al. 2009, \apj,
706, L138

\bibitem[Abdo et al.(2010)]{grb090510_Fermi} Abdo, A.~A. et al. 2010,
\apj, submitted

\bibitem[Asano et al.(2009)]{agm09} Asano, K., Guiriec, S., \&
  M{\'e}sz{\'a}ros, P.\ 2009, \apjl, 705, L191

\bibitem[Atwood et al.(2009)]{LAT} Atwood, W.~B., et al. 2009, \apjs,
697, 1071

\bibitem[Blandford \& McKee(1976)]{bm76} Blandford, R.~D., \& McKee,
C.~F.\ 1976, Physics of Fluids, 19, 1130

\bibitem[B\"ottcher \& Dermer(1998)]{bd98} B\"ottcher, M., \& Dermer,
C.~D.\ 1998, \apjl, 499, L131

\bibitem[Cenko et al.(2010)]{c10} Cenko, S.~B., et al.\ 
2010, arXiv:1004.2900


\bibitem[Chevalier \& Li(2000)]{cl00} Chevalier, R.~A. \& Li,
Z.-Y. 2000, \apj, 536, 195

\bibitem[Corsi et al.(2009)]{cgp09} Corsi, A., Guetta, D., \& Piro,
L.\ 2009, arXiv:0911.4453

\bibitem[De Pasquale et al.(2010)]{grb090510_afterglow} De Pasquale,
M., et al.\ 2010, \apjl, 709, L146

\bibitem[Dermer \& Mitman(1999)]{dm99} Dermer, C.~D., \& Mitman,
K.~E.\ 1999, \apjl, 513, L5


\bibitem[Dermer et al.(2000)]{dcm00} Dermer, C.~D., Chiang, 
J., \& Mitman, K.~E.\ 2000, \apj, 537, 785

\bibitem[Eichler \& Jontof-Hutter(2005)]{ej05} Eichler, D. \&
Jontof-Hutter, D. 2005, \apj, 635, 1182

\bibitem[Freedman \& Waxman(2001)]{fw01} Freedman, D.~L. \& Waxman,
E. 2001, \apj, 547, 922

\bibitem[Gao et al.(2009)]{gmxf09} Gao, W.-H., Mao, J., Xu, 
D., \& Fan, Y.-Z.\ 2009, \apjl, 706, L33

\bibitem[Gehrels et al.(2004)]{Swift} Gehrels, N., et al.\ 
2004, \apj, 611, 1005

\bibitem[Finke et al.(2010)]{frd10} Finke, J.~D., Razzaque, 
S., \& Dermer, C.~D.\ 2010, \apj, 712, 238

\bibitem[Ghirlanda et al.(2009)]{ggn09} Ghirlanda, G., Ghisellini,
G., \& Nava, L.\ 2009, arXiv:0909.0016

\bibitem[Ghisellini et al.(2009)]{ggnc09} Ghisellini, G., Ghirlanda,
G., Nava, L., \& Celotti, A.\ 2009, arXiv:0910.2459

\bibitem[Granot \& Sari(2002)]{gs02} Granot, J., \& Sari, R.\ 2002,
\apj, 568, 820

\bibitem[Hoversten et al.(2009)]{grb090510_Swift} Hoversten, E.~A., 
Krimm, H.~A., Grupe, D., Kuin, N.~P.~M., Barthelmy, S.~D., Burrows, D.~N., 
Roming, P., \& Gehrels, N.\ 2009, GCN Report, 218, 1

\bibitem[Hurley et al.(1994)]{hur94} Hurley, K., et al.\
1994, \nat, 372, 652

\bibitem[Kumar \& Barniol Duran(2009a)]{kb09a} Kumar, P., \& Barniol
  Duran, R.\ 2009, \mnras, L340

\bibitem[Kumar \& Barniol Duran(2009b)]{kb09b} Kumar, P., \& Barniol
  Duran, R.\ 2009, arXiv:0910.5726

\bibitem[Meegan et al.(2009)]{GBM} Meegan, C., et al.\ 2009, \apj,
702, 791

\bibitem[M{\'e}sz{\'a}ros \& Rees(1997)]{mr97} M{\'e}sz{\'a}ros, P.,
\& Rees, M.~J.\ 1997, \apj, 476, 232

\bibitem[Nakar(2007)]{n07} Nakar, E.\ 2007, \physrep, 442, 
166

\bibitem[Panaitescu \& Kumar(2000)]{pk00} Panaitescu, A. \& Kumar,
P. 2000, \apj, 543, 66

\bibitem[Rau et al.(2009)]{grb090510_redshift} Rau, A., McBreen, S., 
\& Kruehler, T.\ 2009, GRB Coordinates Network, 9353, 1

\bibitem[Razzaque et al.(2004)]{rmz04} Razzaque, S., M{\'e}sz{\'a}ros,
P., \& Zhang, B.\ 2004, \apj, 613, 1072

\bibitem[Razzaque \& M{\'e}sz{\'a}ros(2006)]{rm06} Razzaque, S., \&
M{\'e}sz{\'a}ros, P.\ 2006, \apj, 650, 998

\bibitem[Razzaque et al.(2010)]{rdf10} Razzaque, S., Dermer, C.~D., \&
Finke, J.~D.\ 2010, The Open Astronomy Journal (accepted)
arXiv:0908.0513

\bibitem[Razzaque et al.(2009)]{rdf09} Razzaque, S., Dermer, C.~D., \&
Finke, J.~D.\ 2009, \apj, 697, 483

\bibitem[Sari(1997)]{sari97} Sari, R. 1997, \apjl, 489, L37

\bibitem[Sari et al.(1998)]{spn98} Sari, R., Piran, T. \& Narayan,
R. 1998, \apjl, 497, L17

\bibitem[Sari et al.(1999)]{sph99} Sari, R., Piran, T. \& Halpern,
J.~P. 1999, \apjl, 519, L17

\bibitem[Sari \& Esin(2001)]{se01} Sari, R., \& Esin, A.~A.\ 2001,
\apj, 548, 787

\bibitem[Toma et al.(2010)]{twm10} Toma, K., Wu, X.-F., 
\& Meszaros, P.\ 2010, arXiv:1002.2634

\bibitem[Totani(1998a)]{tot98a} Totani, T.\ 1998a, \apjl, 502, L13

\bibitem[Wang et al.(2009)]{wan09} Wang, X.-Y., Li, Z., Dai, Z.-G., \&
M{\'e}sz{\'a}ros, P.\ 2009, \apjl, 698, L98

\bibitem[Zhang \& M\'esz\'aros(2001)]{zm01} Zhang, B., \&
M\'esz\'aros, P.\ 2001, \apj, 559, 110

\end{thebibliography}
\end{document}